\documentstyle[12pt,epsfig]{article}

\begin{document}

\begin{center}

{\Large\bf An origin of the universe:\\
a model alternative to Big Bang}

\vspace*{1cm}

{\large
Andrzej Mercik and
Szymon Mercik$^{\ a,}$\footnote{Corresponding author. E-mail: szymon.mercik@pwr.wroc.pl}
}\\

\vspace*{2mm}

$^{(a)}$ Institute of Physics,\\
Wroclaw University of Technology,\\
50-370 Wroclaw, Poland\\

\end{center}

\vspace*{1cm}

\noindent

{\bf Abstract:} We propose a new approach to the model of an origin of the universe built by Oscar Klein and Hannes Alfv\'{e}n. Some modifications of assumptions underlying the model result in a possible scenario of the universe creation consistent with observations. We explain the large scale structre of the universe and we estimate the Hubble constant value as well as the number of galaxies in the universe. The model does not require many assumptions made in the model based on the Big Bang idea.\\

\noindent



\vspace*{1cm}

\section{Introduction}

The Big Bang model has been considered to be the only currently valid one for many years. However, there are some astronomical observations not explained well enough what keeps forcing the Big Bang supporters to introduce new assumptions. For instance, the homogeneity
and isotropy of the universe on large scales ($>$~100~Mpc), its spatial flatness, and also the distribution of galaxies as well as the fluctuations in the cosmic microwave background is explained by inflation proposed by Alan~H.~Guth~\cite{Guth}. The discrepancy between the predicted and estimated universe matter density is eliminated by the assumption on an existence of dark barion matter, exotic particles (WIMP), and dark energy~\cite{Jungman}. The increasing universe expansion rate is supposed to be covered by a non-zero value of the cosmological constance~\cite{Cohn,Riess}. Also the problem of the initial conditions of the universe resulted in many very abstract hypothesis not supported by experimentally verified physical theories~\cite{Carroll,Gasperini,Steinhardt,Wang}.

Unfortunately, the broad approval of the Big Bang model and its further development caused that alternative models sank into oblivion. In particular, the model built by Oscar Klein and Hannes Alfv\'{e}n~\cite{Alfven} in the 50-ties and 60-ties of last century almost passed out of memory. Their model assumed that the universe evolved from an ambiplsma cloud, i.e. a mixture of protons and antiprotons (heavy ambiplasma) as well as negatons and positons (light ambiplasma). Despite the idea in a natural way explained the universe expansion it did not gain broad support and finally was rejected. It was because the model was based on an assumption that the matter-antimatter symmetry specific for the elementary particles scale is valid also for the universe itself while none of astronomical observations indicated a large scale antimatter clusters. Moreover, from this point of view the universe is extremely asymmetric and antimatter exists only in trace amount.

We present a new approach to the model built by Klein and Alfv\'{e}n by introducing few modifications making the model much better fitting the experimental and observational facts. The main change is a rejection of the original assumption on the perfect matter-antimatter symmetry. It results in an alternative to the Big Bang model explaining our universe evolution without many additional assumptions.

\section{Creation of the observed universe}

To start building the universe model one should accept an assumption on existence of a {\it mother space}. The space is three-dimensional and Euklidesian at large scale but locally the space geometry can differ from Euklidesian as well as an energetic fluctuations can occur. The physics laws result from the {\it mother space} properties.

\subsection{An ambiplasma cloud creation}

As a consequence of energy fluctuations and high energy photons collisions matter-antimatter particle pairs are created. Unstable particles disintegrate and, finally, only few kinds of particles survive: photons, neutrinos and antineutrinos, negatons and positons, as well as protons and antiprotons. A chaotic movement of the particles results in macroscopic fluctuations, i.e. regions of increased ambiplasma density. Gravity force enhances the spatial accumulations creating ambiplasma clouds which start an independent evolution.

The most probably cloud's composition is balanced -- there should be roughly the same amount of matter and antimatter. For instance, if the probabilities of being an antimatter and a matter particle read $p$ and $q=1-p$ respectively, than having $N$ particles the expected (or the average) number of particles reads $\mu=Np$ and its standard deviation equals to $\sigma=\sqrt{Np(1-p)}$. The relative error reads $\sigma/\mu=\sqrt{(1-p)/(Np)}$ so having $10^{80}$ particles and $p\approx 1/2$ one has an immediate result that the statistical fluctuations are of order of $10^{-40}$ and they are meaningless. Nevertheless, the above result is obtained by an assumption, that the probability of matter ($p$) and antimatter $q=1-p$ particle creation is the same. If one takes into consideration a possible difference between the probabilities the situation changes and a cloud with more matter than antimatter can be created. The difference seems to be not big but from the other hand it is suggested by CP symmetry breaking known from the elementary particles physics. So the statistical fluctuation is still not significant but the average number of particles and antiparticles are different and the difference reads $N(2p-1)$. As a consequence from a large number of created ambiplasma clouds some of them can be unbalanced and contain more matter than antimatter (or vice versa).

Despite of an asymmetric matter-antimatter composition a given cloud has to be electrically (quasi)neutral. If a cloud with a net positive electrical charge was created it would attract more negatons than antiprotons since negatons are more mobile. This means that a cloud with more matter than antimatter (i.e. with more protons) would increase the matter-antimatter asymmetry and a cloud with more antimatter would evolve toward symmetry. Again, if a cloud with a net negative electrical charge was created it would attract more positons than protons since positons are more mobile. As a consequence a cloud with more antimatter than matter (i.e. more antiprotons) would increase the matter-antimatter asymmetry and a cloud with more matter would evolve toward symmetry. Finally one can find both symmetric and asymmetric ambiplasma clouds but both kinds should have zero net charge.

\subsection{Evolution of a cloud}

One should consider three forces acting on a given volume element of the cloud:
\begin{itemize}
\item{gravity: $F_g=-\gamma mn$,}
\item{electrical: $F_e=\pm enE$,}
\item{pressure: $F_p=-kT\frac{dn}{dr}$,}
\end{itemize}
where $\gamma$ is a gravity field intensity, $m$ is a particle mass, $n$ is a number of particles in the volume element, $e$ is an electrical charge, $E$ is an electrical field intensity, $k$ is the Stefan-Boltzmann constant, $T$ is temperature, and $r$ is a distance from the cloud's center. The net force
\begin{equation}\label{force}
F=-\gamma mn \pm enE-kT\frac{dn}{dr}
\end{equation}
results in a particle movement and draws the system to the equilibrium (when $F=0$). The process was analyzed in details by Alfv\'{e}n~\cite{Alfven} and we present only the basic steps.

When the cloud is in thermodynamical equilibrium the equation~(\ref{force}) transforms to
\begin{equation}\label{equil}
\frac{1}{n}\frac{dn}{dr}=-\nu\pm \phi
\end{equation}
where $\nu=\gamma m/(kT)$ and $\phi=eE/(kT)$. The solution of~(\ref{equil}) reads
\begin{equation}\label{solution}
n=\eta e^{(-\nu\pm\phi)r}\ ,
\end{equation}
where $\eta$ denotes the given component density in the could's centre. The cloud has to be electrical neutral both globally as well as locally, so the concentrations of protons $n_p^+$, antiprotons $n_p^-$, negatons $n_e^-$, and positons $n_e^+$ have to fulfill the following condition
\[
n_p^+ + n_e^+=n_p^- + n_e^-.
\]
When one applies the condition $\nu_e<<\nu_p$ (since $m_e<<m_p$) the following estimation of the cloud components concentration can be obtained
\begin{equation}\label{prot}
n_p^+=\eta_p^+ e^{-\nu_pr}\sqrt{\frac{\eta_p^- + \eta_e^-e^{\nu_pr}}{\eta_p^+ + \eta_e^+e^{\nu_pr}}}\ ,
\end{equation}
\begin{equation}\label{aprot}
n_p^-=\eta_p^- e^{-\nu_pr}\sqrt{\frac{\eta_p^+ + \eta_e^+e^{\nu_pr}}{\eta_p^- + \eta_e^-e^{\nu_pr}}}\ ,
\end{equation}
\begin{equation}\label{neg}
n_e^-=\eta_e^- e^{-\nu_er}\sqrt{\frac{\eta_e^+ + \eta_p^+e^{\nu_pr}}{\eta_e^- + \eta_p^-e^{\nu_pr}}}\ ,
\end{equation}
\begin{equation}\label{poz}
n_e^+=\eta_e^+ e^{-\nu_er}\sqrt{\frac{\eta_e^- + \eta_p^-e^{\nu_pr}}{\eta_e^+ + \eta_p^+e^{\nu_pr}}}\ .
\end{equation}

Having the equations three basic cases can be considered.

\subsubsection{Pure matter}

In this case $n_p^+=n_e^-$ and $n_e^+=n_p^-=0$ and the formulas~(\ref{prot}) and~(\ref{neg}) take form
\begin{equation}\label{prot1}
n_p^+=\eta_p^+ e^{-\nu_pr/2}
\end{equation}
and
\begin{equation}\label{neg1}
n_e^-=\eta_e^- e^{-\nu_pr/2}
\end{equation}
respectively. So there is no separation of a heavy and light fraction of matter but an electrical field is created. Its intensity can be estimated by substituting~(\ref{prot1}) or~(\ref{neg1}) to~(\ref{equil})
\[
\frac{1}{\eta_p^+ e^{-\nu_pr/2}}\cdot\frac{d}{dr}\left[\eta_p^+ e^{-\nu_pr/2}\right]=-\nu_p+\phi.
\]
After few simple operations one can get
\[
\frac{1}{\eta_p^+ e^{-\nu_pr/2}}\cdot\eta_p^+ e^{-\nu_pr/2}(-\nu_p/2)=-\nu_p+\phi,
\]
so
\[
\nu_p/2=\phi.
\]
Substituting $\nu_p=\gamma m_p/(kT)$ and $\phi=eE/(kT)$ we get the field intensity
\[
E=\frac{\gamma m_p}{2e}.
\]
The field is called the Rosseland electric field.

Discussion for a pure anti-matter cloud is similar.

\subsubsection{Symmetric ambiplasma cloud}

For a symmetric ambiplasma cloud we have $n_p^+=n_p^-$ and $n_e^+=n_e^+$ and the formulas~(\ref{prot})~--~(\ref{poz}) take form
\[
n_p^+=\eta_p^+ e^{-\nu_pr},
\]
\[
n_e^-=\eta_e^- e^{-\nu_er},
\]
\[
n_p^-=\eta_p^- e^{-\nu_pr},
\]
and
\[
n_e^+=\eta_e^+ e^{-\nu_er}
\]
respectively. From the above it follows that there is no separation of matter and anti-matter, but a light fraction is separated from the heavy one. Regions with the same density have $m_p/m_e=1836$ times different depth. The Electrical field does not exist -- substituting the above equations to~(\ref{equil}) one gets $E=0$.

\subsubsection{Asymmetric ambiplasma -- more matter than anti-matter}

\begin{figure}[htbp]
\centerline{\epsfxsize=12cm \epsfbox{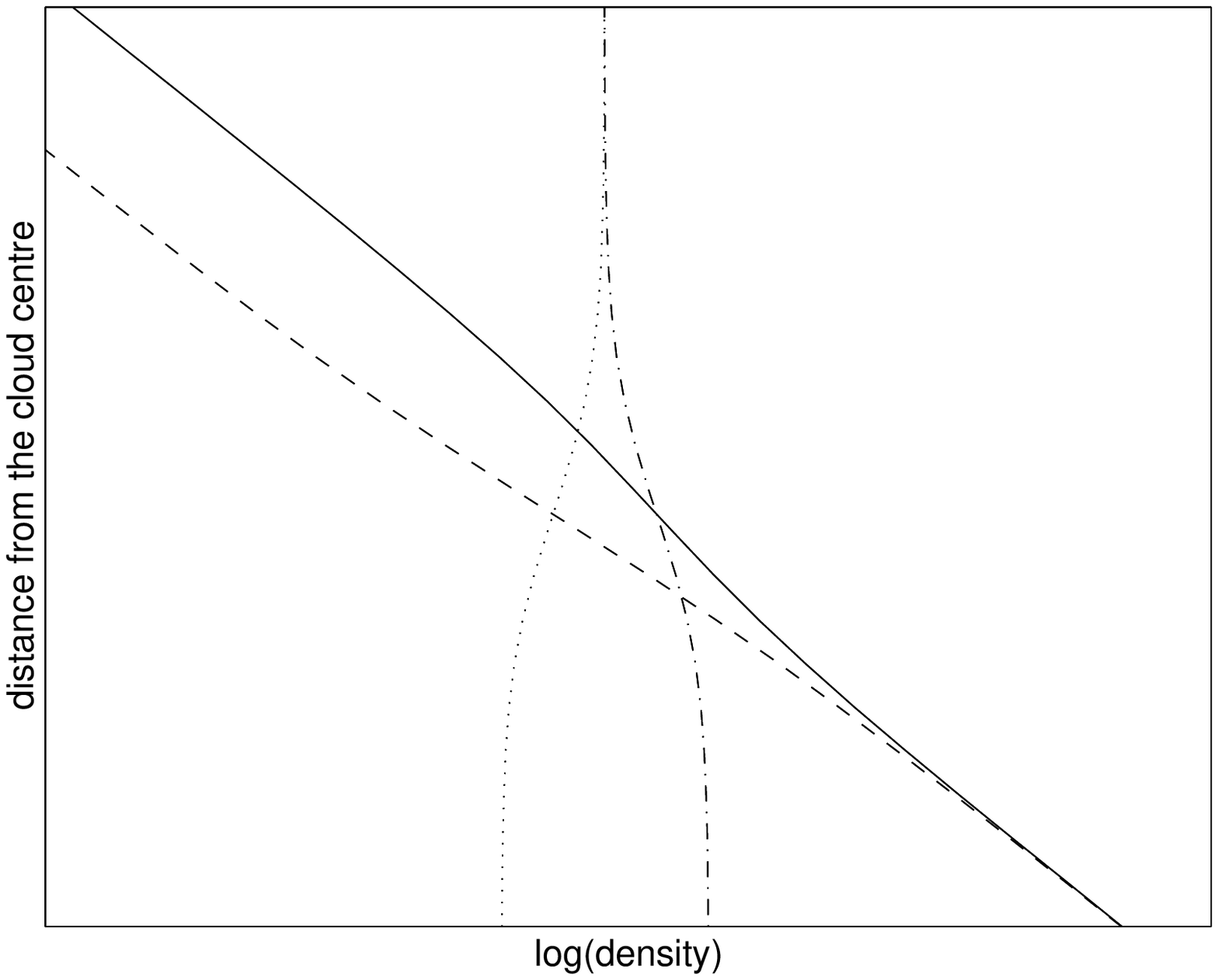}}
\caption{The concentration of matter and anti-matter components as a function of distance from the cloud centre~\cite{Alfven}. The solid line denotes protons, dashed line -- antiprotons, dotted line -- positons, and dashed-dotted line -- electrons.}\label{density_fig}
\end{figure}

In this case the cloud splits into three regions~\cite{Alfven}:
\begin{enumerate}
\item{A central one consisting of symmetric heavy ambiplasma. There is few negatons and even less positons. Electric field intensity is $0$.}
\item{A transitional region with both matter and anti-matter, but the concentration of antiprotons is much less than in the central region and decreases fast towards the surface of the cloud. The negatons' and protons' concentrations are equals but they are bigger than a concentration of positons. The electric field intensity reads $E=\gamma m_p/(2e)$.}
\item{A surface region with symmetric light ambiplasma and no electric field.}
\end{enumerate}
The components concentration dependence on depth is presented in Fig.~\ref{density_fig}. An asymmetric cloud with more anti-matter splits into analogous regions.

Probability of creation of a cloud composed only of matter or anti-matter is so small that the cases can be not considered as a possible way of the universe creation. If one do not exclude a possibility that probabilities of creation of a matter and an antimatter particle are not perfectly the same then also ideally symmetric cloud is not very likely to be created. So the possible universe creation scenario should be developed based on an asymmetric ambiplasma cloud with the three regions described above.

\subsection{Annihilation inside the ambiplasma cloud}

Electrons annihilation forced by weak interaction is very little probably and does not matter for the cloud evolution. As a most probably result of reactions driven by electromagnetic forces two photons are created. In a non-relativistic approach the average life time of a positon in negatons gas is given by Dirac formula
\begin{equation}\label{Dirac}
\tau_e=\frac{k_e}{cn_e^-\sigma_0},
\end{equation}
where $n_e^-$ is negatons' concentration, $k_e\approx 1$ and increases for relativistic velocities. The parameter $\sigma_0$ is given by the following formula
\[
\sigma_0=\pi r_0^2=\frac{e^4}{16\pi\varepsilon_0^2m_e^2c^2} =2.46\cdot10^{-29}\ \rm{[m^2]}
\]
and $r_0$ is the classic electron's radius.

Annihilation of protons is mainly driven by strong interaction and results in pions' creation. An average life time of an antiproton in a gas of protons is given by
\begin{equation}\label{Dirac1}
\tau_p=\frac{k_p}{cn_p^+\sigma_0},
\end{equation}
where $k_p\approx 1$ and is almost constant until velocities reaches relativistic region. Unstable particles created during the protons annihilation disintegrate fast and only few kinds of particle survive: electrons carrying 16\% of the emitted energy, neutrinos (50\% of the energy) and photons (34\% of the energy)~\cite{Wlasow}.

Substitution of the parameters' values to the formulas~(\ref{Dirac}) and~(\ref{Dirac1}) results in the average life time of an antiproton (or a positon) as a function of particles (i.e. protons or electrons) concentration
\[
\tau=\frac{1.33\cdot10^{20}k}{n}\ {\rm[s]}=\frac{4.3\cdot10^{12}k}{n}\ {\rm[year]}.
\]
The power of energy emitted during the annihilation process is given by
\begin{equation}\label{power}
P_e=\frac{2 m_e c^2 N_e^-}{\tau_e}=\frac{8 \pi}{3 k_e}m_e c^3 n_e^2 \sigma_0 r^3,
\end{equation}
where $N_e^-$ denotes the total count of electrons in the cloud. The analogous formula for protons reads
\begin{equation}\label{power1}
P_p=\frac{0.5\cdot 2 m_p c^2 N_p^+}{\tau_p}=\frac{4 \pi}{3 k_p}m_p c^3 n_p^2 \sigma_0 r^3,
\end{equation}
and is based on an assumption that the created positons and negatons also annihilate.

It follows from the equations~(\ref{power}) and~(\ref{power1}) that in a symmetrical ambiplasma cloud the power of radiation emitted as a result of protons annihilation is about 900 times greater than the power generated by electrons annihilation. For this reason we neglect the impact of the electrons annihilation. Assuming $n_pm_p=\rho$ the power of radiation can be given as a function of the cloud density
\begin{equation}\label{power2}
P=\frac{4 \pi}{3 k_p}\cdot\frac{\rho^2 c^3 \sigma_0 r^3}{m_p}.
\end{equation}

In the first stages of the cloud evolution the number of particles and their concentration change very slowly and the annihilation rate is also almost constant. The dependence of the average density of the radiation energy $u$ in the central section of the cloud on time $t$ reads
\[
u=\frac{Pt}{\frac{4}{3}\pi r^3}=\frac{1}{k_p}\cdot\frac{\rho^2 c^3 \sigma_0}{m_p}t
\]
what results in
\[
u \approx 1.3\cdot10^{23}\rho^2t\ \left[\rm{\frac{J}{m^3}}\right].
\]
Combining the value with the Stefan-Boltzmann law
\[
u=\alpha T^4
\]
with $\alpha=7.6\cdot10^{-16}\ \rm{J/m^3K^4}$ one gets a dependence of the cloud temperature on time in the early stages of its evolution
\[
T=\sqrt[4]{1.7\cdot10^{38}\rho^2t}\propto\sqrt[4]{t}.
\]
The temperature is proportional to the 4-th root of the time so when the cloud formulates and its density is small and slowly varying the annihilation almost does not impact the cloud's evolution.

\subsection{Second step of the cloud evolution}

Just after the cloud creates in the {\it mother space} its radiation and electrostatic energies are very small so the main factor of its energy is gravity one given by
\begin{equation}\label{grawitacja}
E_p=-\frac{16}{3} \pi^2 G \rho_0^2 \int_0^r{r^4 dr}.
\end{equation}
Since the gravity energy is negative the cloud starts contracting. The contraction stopps when the radius of the cloud is not smaller than its gravity radius
\begin{equation}\label{grawitacyjny}
r_g=\frac{2 G M_0}{c^2},
\end{equation}
since if the size of the cloud were lower the cloud would collapse. $M_0$ is the initial mass of the cloud. So the upper limit for the cloud density reads
\[
\rho_0=\frac{M_0}{\frac{4}{3}\pi r_g^3}=\frac{3 c^6}{32\pi G^3 M_0^2}.
\]
Integration of the equation~(\ref{grawitacja}) with the upper integration limit~(\ref{grawitacyjny}) and using the above formula for density results in the gravity energy of a cloud which contracts to its gravity radius~(\ref{grawitacyjny})
\[
E_p=-\frac{3}{10} M_0 c^2.
\]
The energy is the minimal one which is necessary to be emitted during the annihilation processes to make the cloud expanding.

During hel synthesis of two protons 0.7\% of the reagents' mass changes into energy. Comparing to the annihilation process the hel synthesis one is much less effective and supplies the cloud with much less energy. For the reason one can assume that the thermonuclear reactions do not impact the cloud evolution if only there is enough of heavy ambiplasma. Half of energy created in positon -- negaton annihilation is carried by neutrinos so it is in practice lost since the particles do not interact with matter. Only the second half of the energy is distributed between electrons, positons, and $\gamma$ quants. In the first step of the cloud evolution the processes run slowly so there is a possibility of annihilation of electrons and positons and an equilibrium between matter and antimatter is reached. Thanks to this matter reaches kinetic energy big enough to initiate the could expansion. To get enough kinetic energy the cloud mass has to be reduced by about 60\% of its initial value.

When the cloud keeps contracting the annihilation rate in its interior increases and as a result temperature and pressure also increase. The conditions are not suitable for its internal structure formulation. Nevertheless, in its external part containing almost pure matter the temperature is lower and decreasing diameter of the cloud is suitable for generation of fluctuations.

The radiation of power $P$ results in momentum transport and its intensity is given by
\[
\frac{dP}{dS\ dT}=\frac{P}{4\pi r^2 c}.
\]
The pressure of the radiation results in a force acting on a unit of the cloud volume
\begin{equation}\label{force1}
f_{\gamma}=\frac{P}{4\pi r^2 c}\xi,
\end{equation}
where $\xi$ is the radiation absorption coefficient
\begin{equation}\label{coef}
\xi=\frac{8}{3}\sigma_0 \zeta n_e.
\end{equation}
$n_e$ denotes electrons concentration and $\zeta$ ranges from $0.1$ to 1 since depending on the radiation energy the interaction with matter is mainly by photoelectrical phenomenon, Compton disperse, or creating of couples negaton-positon.

Combining formulas~(\ref{power2}), (\ref{force1}), and~(\ref{coef}) we get a force coming from the radiation pressure and acting on a unit of the cloud volume distant by $r$ from the cloud centre
\begin{equation}\label{force2}
f_{\gamma}=\frac{8\zeta}{9 k_p}\cdot\frac{c^2 \sigma_0^2 \rho^2 r n_e}{m_p},
\end{equation}
and the gravity force equals to
\begin{equation}\label{force3}
f_g=-\frac{4}{3}\pi G \rho^2 r.
\end{equation}
The radial acceleration of a unit volume of the cloud is given by
\begin{equation}\label{acceleration}
\rho\left\{\frac{dv}{dt} + \frac{dv}{dr}\right\}=f_{\gamma}+f_g.
\end{equation}
A change of the density reads
\begin{equation}\label{density_ch}
\frac{d\rho}{dt}=-3\varsigma\rho - \frac{\pi}{2 k_p}\rho^2,
\end{equation}
where $\varsigma$ is a coefficient in a relationship between a velocity of the volume unit and its distance from the cloud centre
\begin{equation}\label{prop}
v=\varsigma r.
\end{equation}
$\varsigma$ can be estimated from the equation~(\ref{acceleration}) by substituting~(\ref{force2}), (\ref{force3}), and~(\ref{prop})
\[
\frac{d\varsigma}{dt}+\varsigma^2=\frac{4}{3}\pi \left(\frac{2\pi\zeta n_e}{3 k_p} -1 \right).
\]

The delay between the radiation creation and the pressure appearance follows from the limited speed of light and is important for the universe evolution.

\begin{figure}[htbp]
\centerline{\epsfxsize=8cm \epsfbox{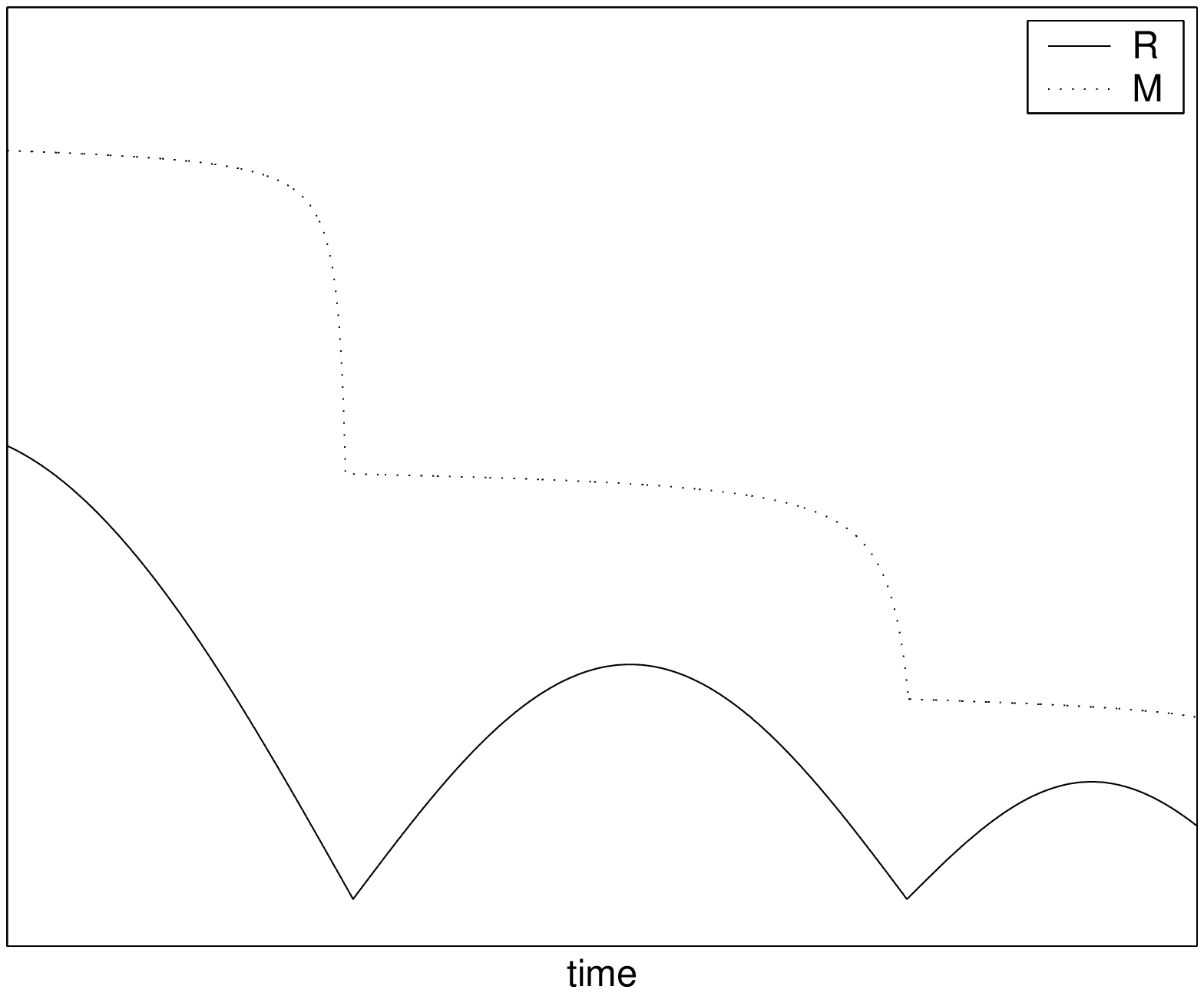} \epsfxsize=8cm \epsfbox{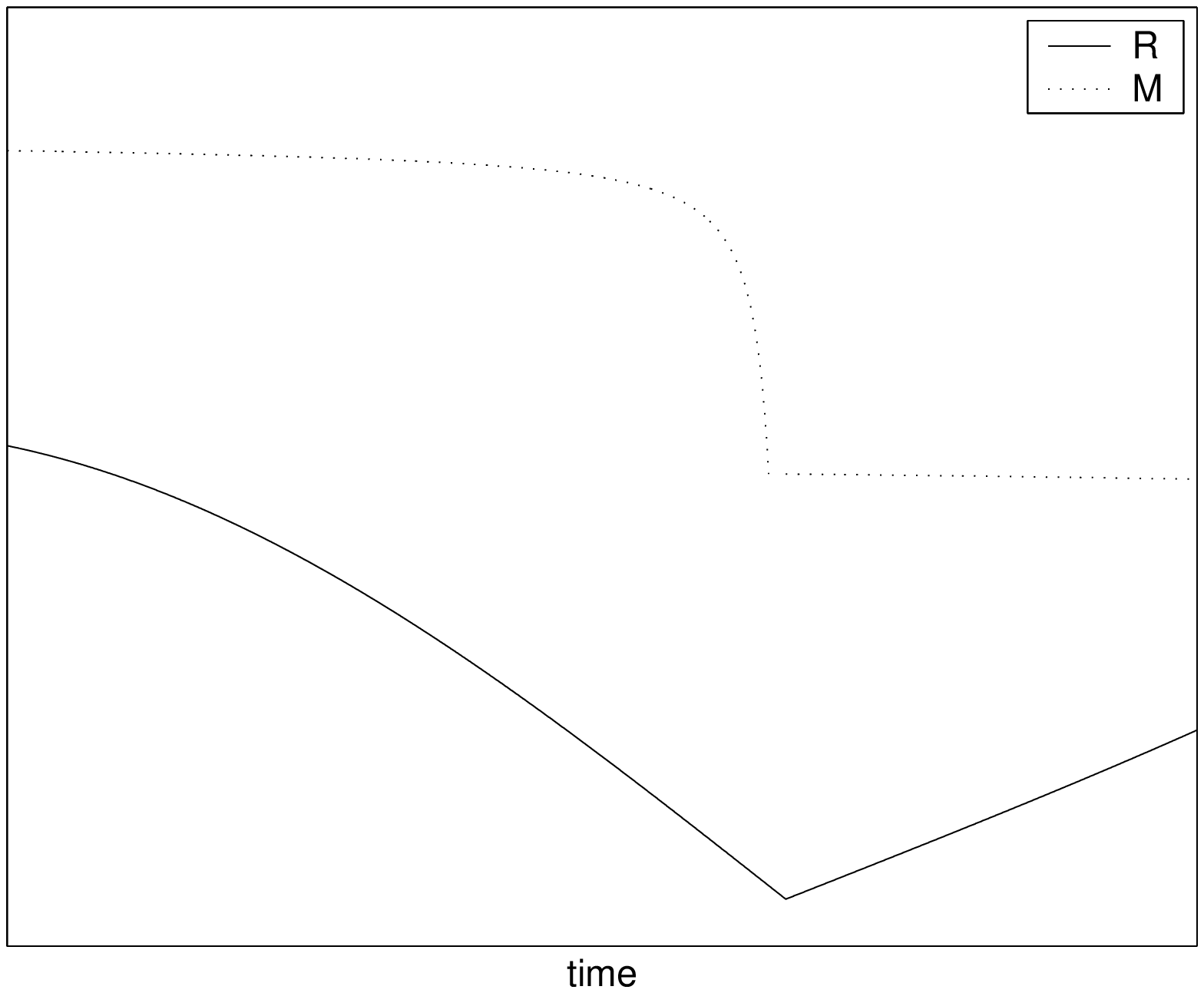}}
\caption{Changes of radius $R$ and mass $M$ of an ambiplasma cloud. Oscillation until the total mass is lost (left) and unlimited expanding after the initial phase of contraction (right)~\cite{Bonnevier}.}\label{osc_fig}
\end{figure}

Bonnevier made an exact analysis of the equations describing the symmetric ambiplasma cloud evolution~\cite{Bonnevier}. He got many different models depending on values of the parameters. The models can be split into two main groups (Fig.~\ref{osc_fig}):
\begin{enumerate}
\item{oscillating until the cloud loses its mass because of its completely annihilation,}
\item{unlimited expanding after the initial phase of contraction.}
\end{enumerate}

In the case of the asymmetric ambiplasma cloud the results mean that in the central section of the cloud the annihilation process continues until the matter and antimatter completely vanish changing into energy. The region can oscillate but can not expand unlimited since it is surrounded by layer of matter. This layer keeps the central section compressed and depending on how much mass the cloud lost in the annihilation process it collapses into black hole or expands.

\subsection{Third step of the cloud evolution}

\begin{figure}[htbp]
\centerline{\epsfxsize=6cm \epsfbox{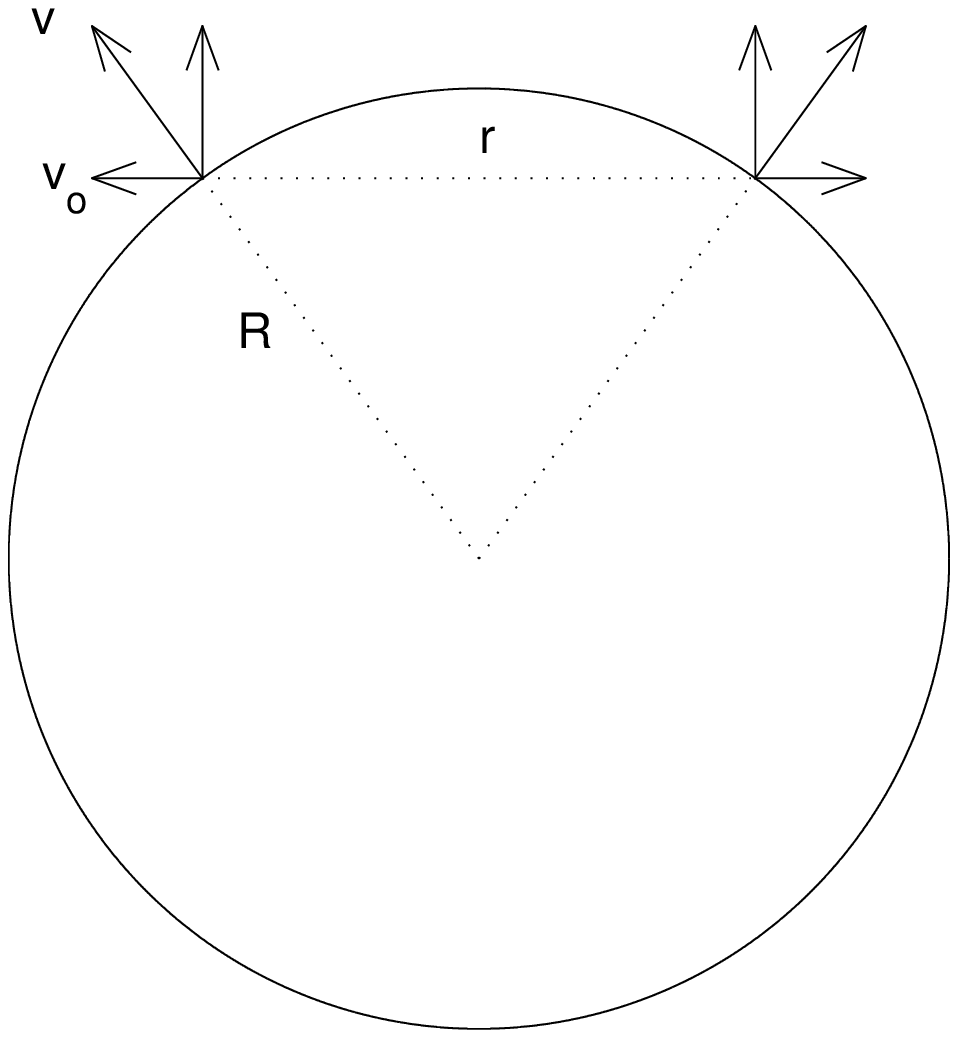}}
\caption{Illustration to the formula~(\ref{Hubble}).}\label{hubble_fig}
\end{figure}

When the cloud contracts its density increases and the annihilation speeds up. As a consequence more radiation is created and its pressure makes the cloud contraction slower. If there is enough energy to stop the contraction the cloud begin to expand. The total cloud energy has to be at least non-negative to initiate expansion. It means that at least 60\% of its initial mass $M_0$ has to be annihilated and the emitted energy has to be absorbed by the layer surrounding the centre. Taking into account that half of the energy is carried out by neutrinos we have that
\begin{equation}\label{equation1}
0.3 M_0 c^2 = 0.4 M_0 c^2 \left(\frac{1}{\sqrt{1-\frac{v^2}{c^2}}}-1 \right).
\end{equation}
From equation~(\ref{equation1}) it follows that the velocity what can be reached by the external layer of the could in this stage of the cloud evolution equals to
\[
v\approx 0.82 c=2.46\cdot 10^{8} \rm{\left[\frac{m}{s}\right]}.
\]
The speed at which a distance between two parts of the external layers increases reads (Fig.~\ref{hubble_fig})
\begin{equation}\label{Hubble0}
v_o=\frac{2 v \frac{r}{2R}}{1+\frac{\left(v\frac{r}{2R}\right)^2}{c^2}}= \frac{\frac{v}{R}r}{1+\frac{v^2 r^2}{4 c^2 R^2}}\ .
\end{equation}
If $v^2 r^2<<4 c^2 R^2$ the formula can be simplified
\begin{equation}\label{Hubble}
v_o=Hr\ .
\end{equation}

The formula~(\ref{Hubble}) represents the Hubble law with $H=v/R$ called the Hubble parameter. Since $v\approx 0.82 c$ the law applies well for $r<<2.5R$.

When the cloud stops contracting and starts expanding the transitional region decreases. It is because of two reasons:
\begin{enumerate}
\item{Gravity force is bigger for the layers more distant from the centre of the cloud, and}
\item{Radiation pressure (so also a force acting on a unit surface) decreases proportional to square of the distance from the centre.}
\end{enumerate}
So the net inward force is bigger for layers more distant from the centre.

The phenomena supports increasing of the fluctuations and since the layer is thinner and thinner the fluctuations should create a set of fibers or sponge-like structures.

When the cloud switches from contraction to expansion there can still be some unannihilated symmetrical ambiplasma in its central region. In this case the radiation is still created in the centre of the cloud and the its power changes periodically as a result of the kernel pulses. So the speed of the external layer expansion can be slowed down and than again accelerated. Another consequence of the central region pulses can be creation of a standing wave. When temperature falls dawn below 3000K and atoms are created the velocity of the wave is given by
\[
v_w=\sqrt{\frac{C_p}{C_v}\cdot\frac{RT}{\mu}},
\]
where $C_p/C_v\approx 1.7$, $R=8.31\ \rm{J/mol\cdot K}$, and $\mu=10^{-3}\ \rm{kg/mol}$. Matter concentrates in knots of the wave in the sphere and the number of the knots reads
\[
n=\frac{2\pi}{v}\cdot\frac{r}{\tau}.
\]
If one assumes that the average velocity of the pulses reads $\hat{v}=2r/\tau$ than the number of the knots equals to
\begin{equation}\label{Galaxies}
n=\pi \hat{v} \sqrt{\frac{C_p}{C_v}\cdot\frac{RT}{\mu}}.
\end{equation}
The number of knots on a surface of a sphere is of order of $n^2$ and defines a number of matter concentrations. The concentrations could be origins of galactics or groups of galactics, analogous to ours Local Group.

The cloud still expands and after a time long enough it spreads out and vanishes in the {\it mother space}.

\section{Observed properties of the universe}

Following the model presented in the previous section we can reconstruct the evolution of the universe. An ambiplasma cloud with zero total electric charge and not symmetrical matter-antimatter composition is created in a {\it mother space} as a result of macroscopic fluctuations. Its initial mass and quantity of matter and antimatter can be estimated using the currently known values of different universe parameters~\cite{Ellis,Freedman}. Observations of a shining or luminous matter indicate that there can be over $10^{78}$ nucleons. Analysis of galaxies dynamics suggest that there can be some five times more of a dark matter which can be detected only by gravity forces. One does not know what is its composition but it can be just barion matter. So an assumption that there are over $10^{79}$ nucleons in the universe seems to be reasonable and as a consequence its mass equals to $M_u=5\cdot 10^{52}\ \rm{kg}$. From the model it follows that it is not more than approx. 40\% of the initial mass of the cloud. So at the moment of its origins the cloud consisted of $10^{80}$ protons and antiprotons and its mass was about $M_{u,0}=10^{53}\ \rm{kg}$. Its radius probably was close to the gravity radius of the cloud and read $r_g=1.5\cdot 10^{26}\ \rm{m}$ and an average density equaled to $\rho_0=10^{-27}\ \rm{[kg/m^3]}$. Of course the numbers are only an estimation but their order of magnitude gives an idea on the initial conditions of the universe evolution.

When the cloud was created gravity forces made it evolving and at the first step of the evolution three layers of the cloud were created: the central one made of heavy ambiplasma of symmetric matter-antimatter composition and of weight of approx. $60\%\cdot M_{u,0}$; a middle layer made of only matter; and the surface layer made of symmetric light ambiplasma. All currently observed objects were created of the matter in the middle layer. The process of fluctuations origination probably started in the contraction phase of the evolution. At the beginning of the expansion phase protogalaxies were created and after this first stars were born.

The oldest observed stars in the universe are aged about $12\cdot 10^{9}\ \rm{years}$ so if one assume that they were created just after the contraction phase switched into the expansion phase one gets that the $R$ parameter in the formula~(\ref{Hubble0}) equals to $R\approx 10^{26}\ \rm{m}$. An analogous result can be obtained if one tries to estimate the $R$ value using the critical density. The density is supposed to read about $10^{-26}\ \rm{[kg/m^3]}$ and the value of an average density is get for a spherical cloud of a mass of $5\cdot 10^{52}\ \rm{kg}$ and a radius $10^{26}\ \rm{m}$. For the value of $R$ the Hubble constant reads
\[
H\approx 2.46\cdot 10^{-18}\ {\rm\left[\frac{1}{s}\right]}\approx 76\ {\rm\left[\frac{km}{s\cdot Mpc}\right]}
\]
so it is consistent with the results got from astrophysical estimations~\cite{Freedman}.

Number of galaxies in the universe can be estimated using the equation~(\ref{Galaxies}). Assuming $\hat{v}=2.46\cdot 10^8\ \rm{m/s}$ and $T=3000\ \rm{K}$ we get
\[
n_g\approx\frac{n^2}{\pi}\approx5\cdot 10^{9}.
\]
Some of the clusters probably become embryos of galaxies and some of them gave birth to groups of galaxies similar to the Local Group. The groups consist of a few dozen of galaxies so the total number of galaxies estimated this way reads approx. $10^{11}$ and it is consistent with observations.

The creation of the matter clusters in the nodes should be reflected in the microwave background radiation. The number of the nodes on a circumference of the matter layer estimated with the equation~(\ref{Galaxies}) reads $n\approx 1.2\cdot 10^{5}$ and an angle between them
\[
\alpha=\frac{2\pi}{n}\approx 5.2\cdot 10^{-5}\ \rm{rad}\approx 0.2\,'\ .
\]
Measurements made by the WMAP satellite prove that the fluctuations are the biggest for angles 100 times bigger~\cite{WMAP}. For the lower angles the fluctuations also exists but their amplitudes are smaller.

\section{Primeval nucleosynthesis}

Contraction of the cloud in the first step if its evolution is faster and faster until gravity force is compensated by the radiation pressure force
\[
\frac{8\zeta}{9 k_p}\cdot\frac{c^2 \sigma_0^2 \rho^2 r n_e}{m_p}-\frac{4}{3}\pi G \rho^2 r=0.
\]
Since in the middle layer made only of matter the count of protons and electrons is (approximately) the same then for the border between the central region and the middle cloud layer we can assume
\[
n_e=n_p=\frac{\rho}{m_p}.
\]
Assuming also that
\[
\frac{8\zeta}{9 k_p}\approx 1
\]
we get an estimation of the matter density at the border between the two regions at the moment when the cloud stops contract and begins its expansion
\[
\rho_1=\frac{4\pi G m_p^2}{3 c^2 \sigma_0^2}=1.5\cdot 10^{-23}\ \rm{\left[\frac{kg}{m^3}\right]}\ .
\]
Comparison of the density value with the value estimated for the initial conditions shows that the density increased $10^4$ times so the size of the cloud decreased about 20 times. The gravity pressure in this region can be estimated as a weight of the matter layer on a unit of surface
\[
p=\rho_1 r_m G \frac{M_w}{r_1^2}\ ,
\]
where $r_m$ is a depth of the matter layer. There are not enough data to estimate the ratio $r_1/r_m$ but if we assume it was of order of 0.1 we get an estimation of a pressure in the bottom of the matter layer of the cloud $p\approx 7\cdot 10^{-7}\ \rm{Pa}$. The temperature in the region estimated by Clapeyron equation gives $T\approx 5\cdot 10^{12}\ \rm{K}$. The further contraction of the cloud still increased its density and temperature.

When temperature is above $9\cdot 10^9\ \rm{K}$ free neutrons are synthesized in the matter layer in a reaction
\[
p+e^- \longrightarrow n + \nu_e -0.8\ \rm{[MeV]}\ .
\]
The reaction run also in the opposite direction resulting in neutrons disintegration so an equilibrium state can be reached between number of protons and neutrons. When the cloud contraction switches into the cloud expansion the matter temperature starts to decrease. It results in faster neutrons disintegration than their synthesis so their concentration begins to decrease. Nevertheless not every neutron vanishes since at the temperature of $10^9\ \rm{K}$ deuter is synthesized and than helium nucleus are created. It is not clear if the proportion between different elements' nucleous concentration is in line with observations~\cite{Olive} and answering the question requires further studies on the model.

\section{Conclusion}

The presented model can not be called a complete one but it shows an interesting way competitive to the Big Bang model. Even if the model is difficult to be accepted since the Big Bang model is considered to be the currently valid one its consideration can open a door to solution of many problems existing in the Big Band model. Moreover, the presented model has many advantages comparing to the Big Bang model:
\begin{enumerate}
\item{Basing on experimental verified physical laws it explains the universe expansion.}
\item{It omits the problem with the original singularity.}
\item{It does not require an assumption about existence of inflation.}
\item{It explains origins of the background radiation and its power distribution.}
\item{It explains big scale universe structure consistent with observations.}
\item{It does not require assumptions on exotic matter components and dark energy.}
\item{It shows that even before first stars were born in the first stage of the universe evolution there had existed conditions necessary for complex atomic nucleus creation.}
\end{enumerate}

For this reason the model deserves to be analyzed and to be considered as a possible walk around of many existing problems with the universe evolution.

\section{Acknowledgements}

The Authors wish to thank to Anna M. Tokarska and Artur Tokarski for a help in the paper preparation.

\begin{thebibliography}{99}
{\small
\bibitem{Alfven} H.~Alfv\'{e}n, Reviews of Modern Physics {\bf 37}, 652 (1965).
\bibitem{Bonnevier} B.~Bonnevier, Ark.~Fys. {\bf 27}, 310 (1964).
\bibitem{Carroll} S.~M.~Carroll, V.~Duvvuri, M.~Trodden, M.~S.~Turner, Phys.~Rev.~D {\bf 70}, 043528 (2004); astro-ph/0306438.
\bibitem{Cohen} A.~G.~Cohen, http://www.slac.stanford.edu/gen/meeting/ssi/1999/media/cohen.pdf
\bibitem{Cohn} J.~D.~Cohn, Astrophys.~J.~Suppl. {\bf 259}, 213 (1998); astro-ph/9807128.
\bibitem{Ellis}	R.~Ellis; astro-ph/0102056.
\bibitem{Freedman} W.~L.~Freedman, J.~R.~Mould, R.~C.~Kennicutt Jr., B.~F.~Madore; astro-ph/9801080.
\bibitem{Gasperini} M.~Gasperini, G.~Veneziano, Phys.~Rept. {\bf 373}, 1 (2003); hep-th/0207130.
\bibitem{Guth} A.~H.~Guth, {\it The Inflationary Universe, The Quest for a New Theory of Cosmic Origins}, (Perseus Books, Reading, Mass., 1998).
\bibitem{Jungman} G.~Jungman, M.~Kamionkowski, K.~Griest, Phys.~Rep. {\bf 267}, 195 (1996); hep-ph/9506380
\bibitem{Olive} K.~A.~Olive, G.~Steigman, T.~P.~Walker; astro-ph/9905320.
\bibitem{Quinn} H.~R.~Quinn, SLAC-PUB-8784, (2001).
\bibitem{Riess} A.~G.~Riess, A.~V.~Filippenko, P.~Challis, A.~Clocchiattia, A.~Diercks, P.~M.~Garnavich, R.~L.~Gilliland, C.~J.~Hogan, S.~Jha, R.~P.~Kirshner, B.~Leibundgut, M.~M.~Phillips, D.~Reiss, B.~P.~Schmidt, R.~A.~Schommer, R.~Ch.~Smith, J.~Spyromilio, Ch.~Stubbs, N.~B.~Suntzeff, J.~Tonry, Astron.~J. {\bf 116}, 1009 (1998); astro-ph/9805201.
\bibitem{Steinhardt} P.~J.~Steinhardt, N.~Turok, Science {\bf 296}, 5572 (2002); hep-th/0111030.
\bibitem{Wang} L.~Wang, R.~R.~Caldwell, I.~P.~Ostriker, P.~J.~Steinhardt, Astrophys.~J. {\bf 530}, 17 (2000); astro-ph/9901388.
\bibitem{Wlasow} N.~A.~Wlasow, Antiwiesciestwo, (Atomizdat, Moscow, 1966).
\bibitem{WMAP} Results of WMAP satelite observations (map.gsfc.nasa.gov).
}
\end{thebibliography}
\end{document}